\documentclass[12pt,preprint]{aastex}

\usepackage{amsmath}
\usepackage{graphicx}
\usepackage{comment}
\usepackage{color} 
\newcommand{\fig}[1]{Fig.~\ref{#1}}
\newcommand{\tbl}[1]{Table~\ref{#1}}
\newcommand{\sect}[1]{Section~\ref{#1}}

\begin{document}
\title{Upward and downward catastrophes of coronal magnetic flux ropes in quadrupolar magnetic fields}
\author{Quanhao Zhang\altaffilmark{1}, Yuming Wang\altaffilmark{1,2}, Youqiu Hu\altaffilmark{1}, Rui Liu\altaffilmark{1,3}, Kai Liu\altaffilmark{1,4}, Jiajia Liu\altaffilmark{1,4}}
\altaffiltext{1}{CAS Key Laboratory of Geospace Environment, Department of Geophysics and Planetary Sciences, University of Science and Technology of China, Hefei 230026, China}
\altaffiltext{2}{Synergetic Innovation Center of Quantum Information \& Quantum Physics, University of Science and Technology of China, Hefei, Anhui 230026, China}
\altaffiltext{3}{Collaborative Innovation Center of Astronautical Science and Technology, China}
\altaffiltext{4}{Mengcheng National Geophysical Observatory, School of Earth and Space Sciences, University of Science and Technology of China, Hefei 230026, China}
\email{zhangqh@mail.ustc.edu.cn}

\begin{abstract}
Coronal magnetic flux ropes are closely related to large-scale solar activities. Using a 2.5-dimensional time-dependent ideal magnetohydrodynamic (MHD) model in Cartesian coordinates, we carry out numerical simulations to investigate the evolution of a magnetic system consisting of a flux rope embedded in a fully-closed quadrupolar magnetic field with different photospheric flux distributions. It is found that, when the photospheric flux is not concentrated too much towards the polarity inversion line (PIL) and the constraint exerted by the background field is not too weak, the equilibrium states of the system are divided into two branches: the rope sticks to the photosphere for the lower branch and levitates in the corona for the upper branch. These two branches are connected by an upward catastrophe (from the lower branch to the upper) and a downward catastrophe (from the upper branch to the lower). Our simulations reveal that there exist both upward and downward catastrophes in quadrupolar fields, which may be either force-free or non-force-free. The existence and the properties of these two catastrophes are influenced by the photospheric flux distribution, and a downward catastrophe is always paired with an upward catastrophe. Comparing the decay indices in catastrophic and non-catastrophic cases, we infer that torus unstable might be a necessary but not sufficient condition for a catastrophic system.
\end{abstract}


\section{Introduction}
\label{sec:introduction}
\begin{figure*}
\includegraphics[width=\hsize]{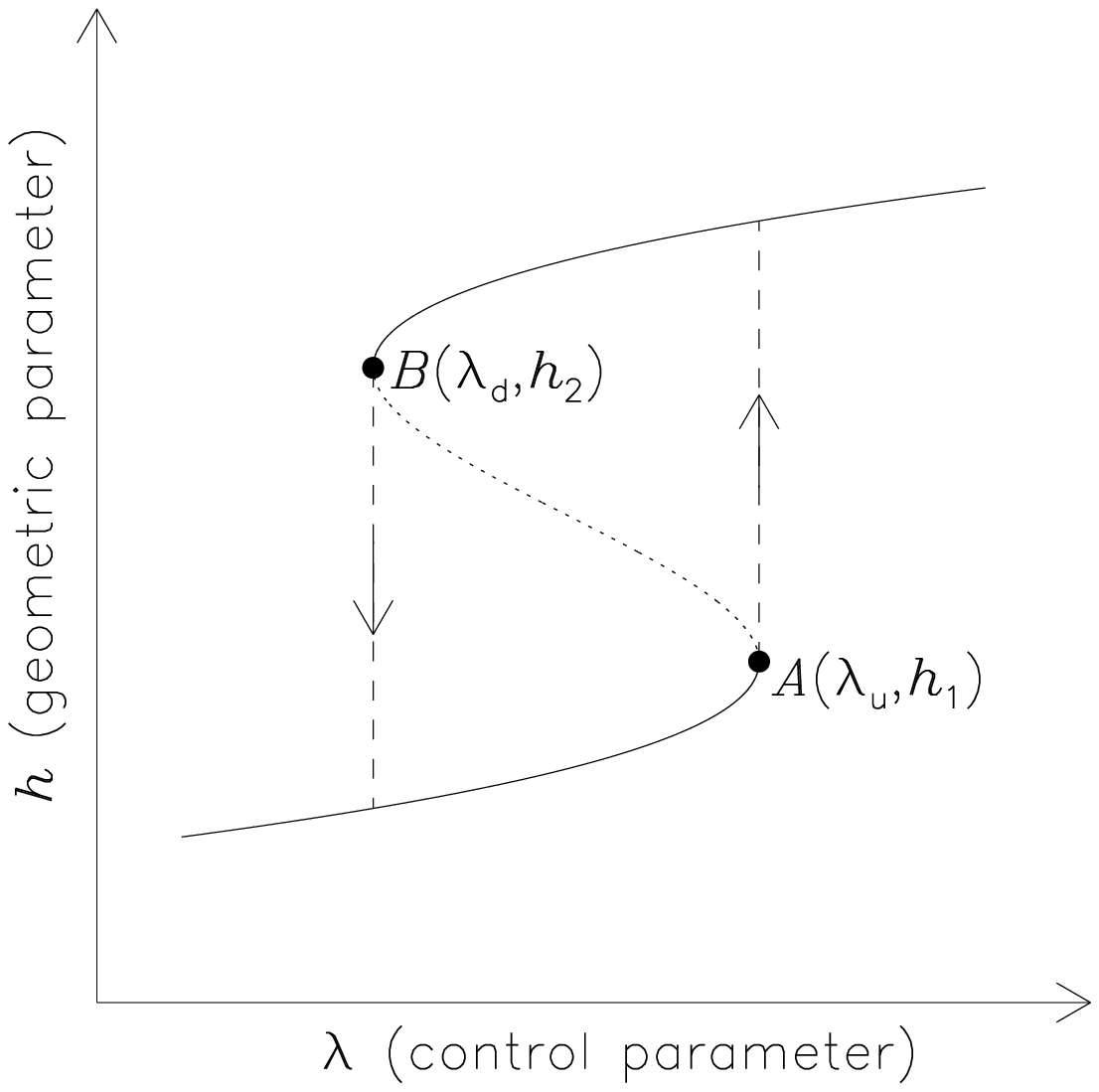}
\caption{Schematic cartoon of the equilibrium manifold with both upward and downward catastrophe. $\lambda$ is the control parameter and h is the geometric parameter. Upward and downward catastrophes occurs at the nose point A and B, respectively, and $\lambda_u$ and $\lambda_d$ are the corresponding catastrophic points.}\label{fig:cart}
\end{figure*}

It is well known that large-scale solar eruptive activities, including prominence/filament eruptions, flares, and coronal mass ejections (CMEs), are closely related to solar magnetic flux ropes \citep[e.g.,][]{Low1996a,Shibata2011a,Chen2011a,Wang2015a}. In order to understand the physical processes of solar eruptive activities, various theoretical models have been proposed to describe the eruption of solar magnetic flux ropes, invoking distinctive physical mechanisms, e.g., magnetic reconnections \citep{Antiochos1999a,Chen2000a,Moore2001a}, MHD instabilities \citep{Amari2000a,Liu2007a,Kliem2006a} and catastrophe. 
Catastrophe of flux rope systems was first proposed by \cite{vanTend1978a}, who concluded that, if the current in a filament exceeds a critical value, a catastrophic loss of equilibrium occurs in the magnetic system. Both analytical and numerical analyses have been made to investigate the catastrophic behaviors of solar magnetic flux ropes, and leading to a common conclusion that catastrophe could be responsible for flux rope eruptions \citep{Forbes1990a,Priest1990a,Forbes1991a,Isenberg1993a,Forbes1995a,Lin2000a,Lin2002b,Zhang2007a,Su2011a}. These studies focus on the equilibrium manifold in parameter space, i.e. the evolution of the equilibrium states of the system as a function of a certain control parameter characterising the physical properties of the system. In analytical analyses, the equilibrium manifold is obtained by solving the force balance equation, whereas in numerical simulations, it is obtained by calculating the different equilibrium states with different values of the control parameter. The critical value of the control parameter at which catastrophe occurs is called catastrophic point, which usually appears as an end or nose point of the equilibrium manifold \citep[][ also see \fig{fig:cart}]{Kliem2014a}. Magnetic free energy is always released and converted to kinetic and thermal energy during catastrophe \citep{Chen2007a}, via both magnetic reconnection and the work done by Lorentz force \citep{Zhang2016a}. Previous studies also demonstrated that catastrophe and instability are intimately related in the evolution of magnetic flux rope systems \citep{Demoulin2010a,Kliem2014a,Longcope2014a}.
\par
Cartesian coordinates are widely used to investigate active region activities. Using a 2.5-dimensional ideal MHD model in Cartesian coordinates, \cite{Hu2001a} found that the equilibrium states of the magnetic system consisting of a flux rope embedded in a partially open bipolar background field are divided into upper and lower branches, and there exists an upward catastrophe from the lower branch to the upper branch. By simulating the evolution of the similar system under different photospheric magnetic conditions, \cite{Zhang2017a} found that the upward catastrophic behavior of the magnetic system is influenced by photospheric magnetic condition, namely, the transition from the equilibrium state with the flux rope sticking to the photosphere (hereafter, ``sticky" state) to that with the flux rope levitating in the corona (hereafter, ``levitating" state) varies with the photospheric magnetic flux distribution. When the photospheric flux is not concentrated too much toward the PIL and the source regions of the bipolar field are not to weak, sticky and levitating states are separated, and correspond to the lower and upper branches, respectively. Otherwise, the transition between the sticky and levitating states is continuous, implying that the system is non-catastrophic. It should be noted that previous studies demonstrated that there are two type of flux rope topologies \citep[e.g.,][]{Savcheva2009a,Green2009a}: the flux rope with its underside rooted in the dense lower atmosphere is called bald-patch separatrix surface (BPSS) configuration \citep{Titov1993a,Titov1999a,Green2011a}, and the flux rope levitating in the corona with a magnetic X-type structure beneath is called hyperbolic flux tube (HFT) configuration \citep{Titov2002a}. These two types of configurations and the transition between them have been observed in many studies based on non-linear force-free field (NLFFF) extrapolations \citep[e.g.,][]{Zhao2014a,Savcheva2015a,Savcheva2016a,Janvier2016a}. Comparing with the sticky state and levitating state, we may infer that the sticky state is consistent with BPSS configurations, and the levitating state, in which the flux rope levitating in the corona with a current sheet below the rope, corresponds to the configuration after the X-type structure in HFT has evolved into a current sheet.
\par
Recently, by expanding the 2.5-dimensional ideal MHD model used in \cite{Hu2001a} to force-free approximations, \cite{Zhang2016a} found that, apart from the well-known upward catastrophe, there also exists a downward catastrophe from the upper branch to the lower branch in a partially open bipolar background field. Just like that happens during upward catastrophe, magnetic energy is also released during downward catastrophe, indicating that downward catastrophe might be a possible mechanism for non-eruptive but energetic activities (e.g., confined flares). \fig{fig:cart} is a schematic cartoon of the equilibrium manifold consisting of both upward and downward catastrophes. Here $\lambda$ is the control parameter and $h$ is the geometric parameter describing the evolution of the equilibrium states (e.g., the height of the rope axis). There are two nose points: A and B, at which upward and downward catastrophe occur, respectively. Obviously, the ill-behaviors of the upward and downward catastrophes result from that the equilibrium manifold is multi-valued within the upward ($\lambda_u$) and downward ($\lambda_d$) catastrophic points.
\par
All the previous numerical simulations about upward and downward catastrophes using the 2.5-dimensional ideal MHD model in Cartesian coordinates are made to investigate the magnetic system in bipolar background fields. Since the magnetic configuration in strong active regions are usually very complex \citep[e.g.,][]{Schrijver2011,Sun2012a,vanDriel2015a}, a quadrupolar background field should be more suitable for analyses of large-scale activities in active region. In this paper, we simulate the evolution of a magnetic system consisting of a flux rope in a fully closed quadrupolar background field, so as to investigate whether upward and downward catastrophes also exist in quadrupolar fields. Moreover, since only the photospheric magnetic conditions can be observed currently, to reveal the influence of the photospheric conditions on catastrophes could shed light on the physical processes of different solar activities. As mentioned above, previous studies have found that the existence and properties of upward catastrophe are affected by photospheric flux distribution. Thus, another intention of this paper is to investigate the influence of photospheric magnetic conditions on downward catastrophe. The sections are arranged as follows: the simulation model in quadrupolar field is introduced in \sect{sec:equation}; the evolutions of the magnetic system with different photospheric magnetic conditions under force-free and non-force-free conditions are demonstrated in \sect{sec:result}; the relationship between catastrophe and torus instability is investigated in \sect{sec:torus}. Finally, a discussion is given in \sect{sec:discuss}.

\section{Basic equations and simulating procedures}
\label{sec:equation}
As mentioned in \sect{sec:introduction}, a Cartesian coordinate system is used here. A magnetic flux function $\psi$ is introduced to denote the magnetic field as follows:
\begin{align}
\textbf{B}=\triangledown\times(\psi\hat{\textbf{\emph{z}}})+B_z\hat{\textbf{\emph{z}}}.\label{equ:mf}
\end{align}
Neglecting the radiation and heat conduction in the energy equation, the 2.5-D MHD equations can be written in the non-dimensional form as:
\begin{align}
&\frac{\partial\rho}{\partial t}+\triangledown\cdot(\rho\textbf{\emph{v}})=0,\label{equ:cal-st}\\
&\frac{\partial\textbf{\emph{v}}}{\partial t}+\textbf{\emph{v}}\cdot\triangledown\textbf{\emph{v}}+\triangledown T +\frac{T}{\rho}\triangledown\rho+\frac{2}{\rho\beta_0}(\vartriangle\psi\triangledown\psi+B_z\triangledown B_z+\triangledown\psi\times\triangledown B_z)+g\hat{\textbf{\emph{y}}}=0,\\
&\frac{\partial\psi}{\partial t}+\textbf{\emph{v}}\cdot\triangledown\psi=0,\\
&\frac{\partial B_z}{\partial t}+\triangledown\cdot(B_z\textbf{\emph{v}})+(\triangledown\psi\times\triangledown v_z)\cdot\hat{\textbf{\emph{z}}}=0,\\
&\frac{\partial T}{\partial t}+\textbf{\emph{v}}\cdot\triangledown T +(\gamma-1)T\triangledown\cdot\textbf{\emph{v}}=0,\label{equ:cal-en}
\end{align}
where $\rho, \textbf{\emph{v}}, T, \psi$ denote the density, velocity, temperature and magnetic flux function, respectively; the subscript $z$ denotes the $z-$component of the parameters, which are parallel to the axis of the flux rope; $\beta_0=2\mu_0\rho_0RT_0L_0^2/\psi_0^2=0.1$ is the characteristic ratio of the gas pressure to the magnetic pressure, where $\mu_0$ and $R$ are the vacuum magnetic permeability and the gas constant, respectively; $\rho_0=3.34\times10^{-16}\mathrm{~g~cm^{-3}}$, $T_0=10^6\mathrm{~K}$, $L_0=10^9\mathrm{~cm}$, and $\psi_0=3.73\times10^9\mathrm{~Mx~cm^{-1}}$ are the characteristic values of density, temperature, length and magnetic flux function, respectively; $g$ is the normalized gravity. The initial corona is static and isothermal with
\begin{align}
T_c\equiv T(0,x,y)=1\times10^6 ~\mathrm{K},\ \  \rho_c\equiv\rho(0,x,y)=\rho_0\mathrm{e}^{-gy}.
\end{align}
\par
In this paper, we study the evolution of the magnetic system under both force-free and non-force-free conditions. Force-free equilibrium solutions are obtained by a relaxation method: reset the temperature and density in the computational domain to their initial values, so that the pressure gradient force is always balanced everywhere by the gravitational force \citep{Hu2004a}.
\par
The background field is a fully-closed quadrupolar field, which is assumed to be symmetrical relative to $y$-axis. The lower boundary $y=0$ corresponds to the photosphere. There are two pairs of positive and negative magnetic surface charges located at the photosphere. For the inner pair, which is closer to the PIL, the positive charge is located at $y=0$ within $-b<x<-a$ and the negative one within $a<x<b$. For the outer pair, the positive and negative charges are located at $y=0$ within $c<x<d$ and $-d<x<-c$, respectively ($a<b<c<d$). The ratio of the charge density of the inner two charges to that of the outer ones is $\sigma$. By complex variable method, the background magnetic field can be cast in a complex variable form
\begin{align}
f(\omega)\equiv B_x-iB_y=\mathrm{ln}\left(\frac{\omega^2-c^2}{\omega^2-d^2}\right)-\sigma\mathrm{ln}\left(\frac{\omega^2-a^2}{\omega^2-b^2}\right)=
\mathrm{ln}\left[\frac{(\omega^2-c^2)(\omega^2-b^2)^\sigma}{(\omega^2-d^2)(\omega^2-a^2)^\sigma}\right],
\end{align}
where $\omega=x+iy$. The magnetic flux function is then calculated by
\begin{align}
\psi(x,y)=\mathrm{Im}\left\lbrace\int f(\omega)d\omega \right\rbrace;\label{equ:complex}
\end{align}
and the flux function at the photosphere can be derived as
\begin{equation}
 \psi_b = \psi(x,0) = \left\{
              \begin{array}{ll}
              {0}, &{x\leqslant-d}\\
              {\pi(x+d)}, &{-d<x\leqslant-c}\\
              {\pi w}, &{-c<x\leqslant-b}\\
              {\pi w-\sigma\pi(x+b)}, &{-b<x\leqslant-a}\\
              {(1-\sigma)\pi w}, &{-a<x\leqslant a}\\
              {\pi w+\sigma\pi(x-b)}, &{ a<x\leqslant b}\\
              {\pi w}, &{ b<x\leqslant c}\\
              {\pi (d-x)}, &{ c<x\leqslant d}\\
              {0}, &{x>d}
              \end{array}  
         \right.
\end{equation}
where $w=b-a=d-c$ is the width of the charges. Here $a, b, c, d,$ and $\sigma$ characterise the photospheric magnetic flux distribution. With different values of $(a, b, c, d, \sigma)$, we obtain different background magnetic configurations. 
\par
With the initial and boundary conditions, equations (\ref{equ:cal-st}) to (\ref{equ:cal-en}) are solved by the multi-step implicit scheme \citep{Hu1989a} to obtain equilibrium solutions of the magnetic system. The computational domain is taken to be $0\leqslant x\leqslant100\mathrm{~Mm}$, $0\leqslant y\leqslant300\mathrm{~Mm}$; a symmetric condition is used for the left side of the domain ($x=0$). During the simulation, potential field conditions are used at the top ($y=300$ Mm) and right ($x=300$ Mm) boundaries, and the flux function at the lower boundary ($y=0$) is fixed at $\psi_b$. 
\par
Starting from a background magnetic configuration with given values of $(a, b, c, d, \sigma)$, following \cite{Hu2000a} and \cite{Hu2001a}, we let a flux rope emerge from the central region of the base, and then the flux rope sticks to the photosphere, resulting in a magnetic system consisting of a flux rope embedded in a fully closed quadrupolar field, which is the initial state for the given group of $(a, b, c, d, \sigma)$. The magnetic properties of the flux rope are characterised by the axial magnetic flux passing through the cross section of the flux rope, $\Phi_z$, and the poloidal magnetic flux of the rope of per unit length along $z$-direction, $\Phi_p$. Note that $\Phi_p$ is simply the difference in $\psi$ between the axis and the outer boundary of the flux rope, negative for the present case with field lines rotating clockwise in the rope. Here we select $\Phi_z$ as the control parameter. In our simulation, we adjust the values of $\Phi_z$with fixed $\Phi_p$ to calculate different equilibrium solutions of the system, namely, we analyze the equilibrium manifold a function of $\Phi_z$, as described by the geometric parameters of the flux rope, including the height of the rope axis, $H$, and the length of the current sheet below the flux rope, $L_c$. For the sticky state, $L_c$ equals $0$, and for the levitating state, there is a current sheet below the flux rope, so that $L_c$ is finite. For background configurations with different values of $(a, b, c, d, \sigma)$, similar procedures are repeated, so that we obtain the equilibrium manifolds of the flux rope system under different photospheric flux distributions (see \sect{sec:result}).
\par
The equilibrium solution with a certain value of $\Phi_z$ is calculated as: first slowly adjust $\Phi_z$ to the target value, and then let the system relax to equilibrium state, during which $\Phi_z$ is maintained to be conserved at the target value, which is achieved by the same numerical measure as that introduced in \cite{Hu2003a}. 

\section{Simulation results}
\label{sec:result}

\subsection{Force-free condition}
\label{sec:free}
First, we analyze the evolution of the magnetic system under force-free conditions, and the evolution of the system is purely determined by magnetic forces. As mentioned in \sect{sec:equation}, different values of $(a, b, c, d, \sigma)$ correspond to different background field, resulting in different magnetic systems. Here we adjust the distance between the inner pair of the charges, $d_s=2a$, and the strength of them, which is characterised by the value of $\sigma$, to obtain magnetic systems with different photospheric flux distributions. The width of these charges is always fixed at $w=b-a=d-c=20$ Mm, and the distance between the inner and the outer pair of charges is also fixed at $D=c-b=5$ Mm. 
\par
\begin{figure*}
\includegraphics[width=\hsize]{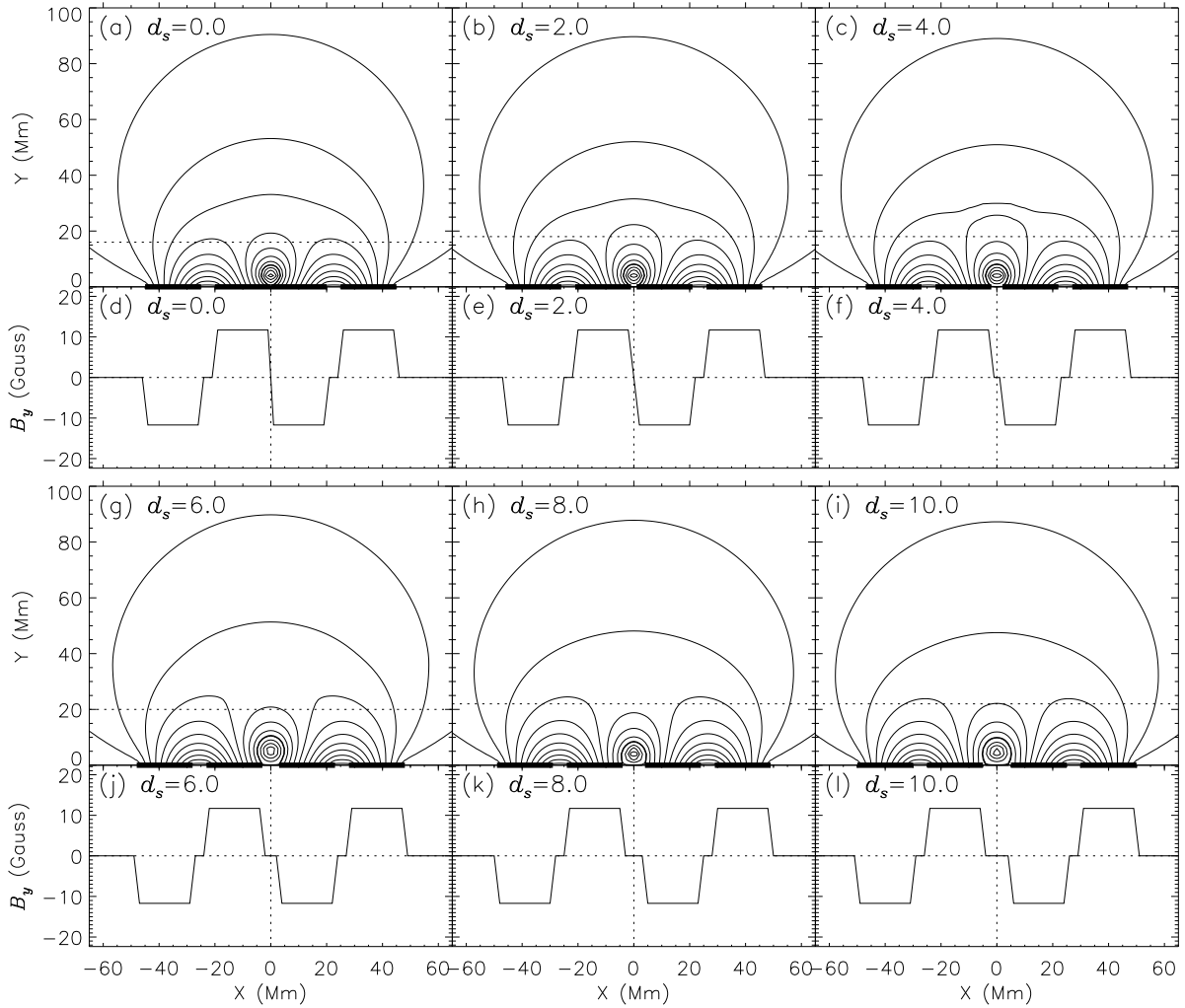}
\caption{The initial configurations and the corresponding normal components of the magnetic field ($B_y$) at the photosphere ($y=0$) for different $d_s$, which is selected to be $0.0, 2.0, 4.0, 6.0, 8.0, 10.0$ Mm, respectively; $\sigma$ is 1.0 for all the six cases. The two pairs of surface magnetic charges for different cases are marked by the black solid lines at $y=0$ in panels (a)-(c) and (g)-(i). The height of the neutral point in the quadrupolar background field is marked by the horizontal dotted line in panels (a)-(c) and (g)-(i). }\label{fig:initd}
\end{figure*}
\begin{figure*}
\includegraphics[width=\hsize]{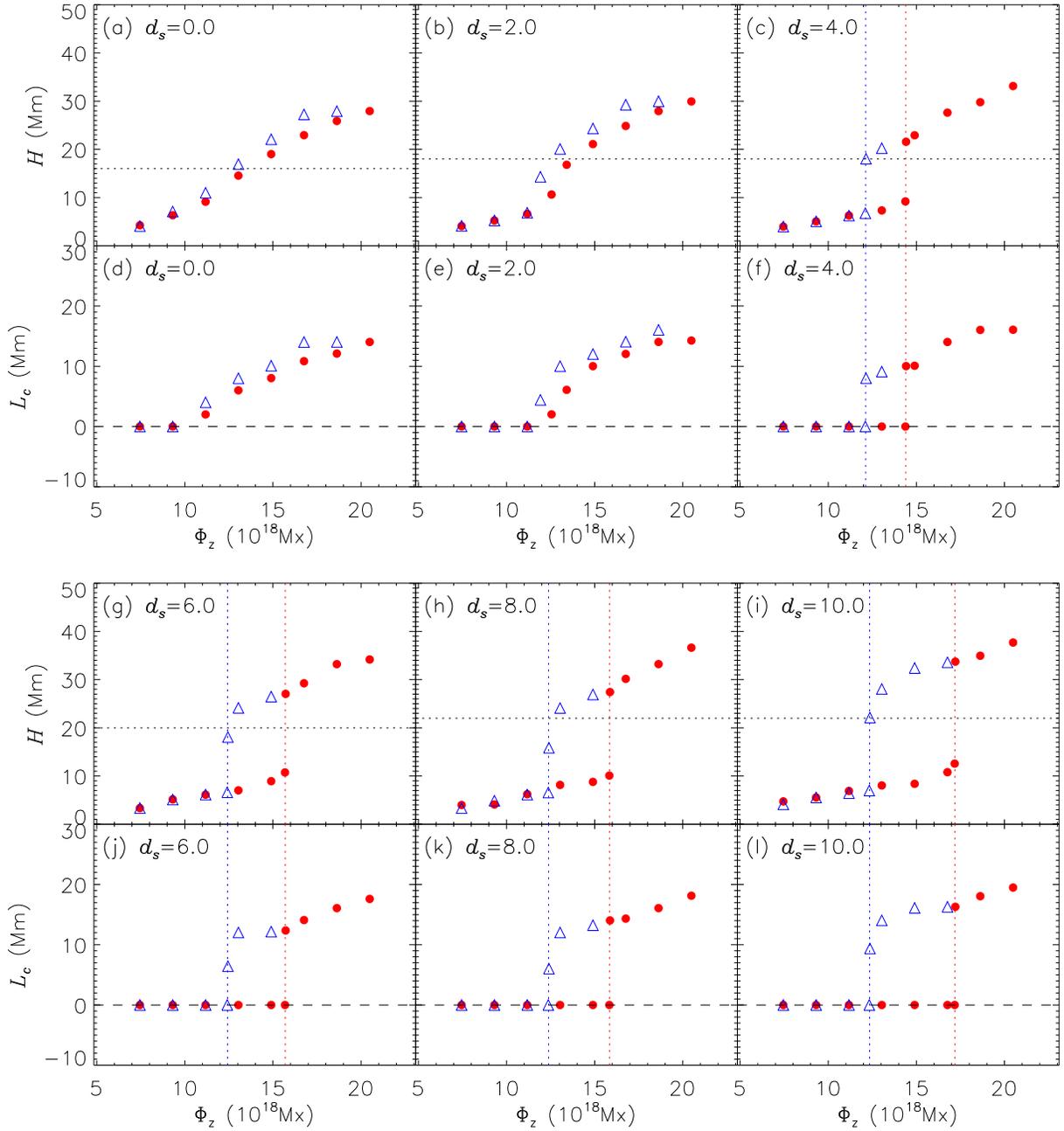}
\caption{The height of the flux rope axis ($H$) and the length of the current sheet below the rope ($L_c$) are shown as functions of the control parameter $\Phi_z$ for quadrupolar background fields with different $d_s$ under force-free conditions. The red points represent the transition from sticky states to levitating states; the blue triangles represent the transition from levitating states to sticky states. The vertical dotted lines represent the catastrophic points of the catastrophic cases. The height of the neutral point in the quadrupolar background field is marked by the horizontal dotted line in panels (a)-(c) and (g)-(i).}\label{fig:fluxd}
\end{figure*}
\begin{figure*}
\includegraphics[width=\hsize]{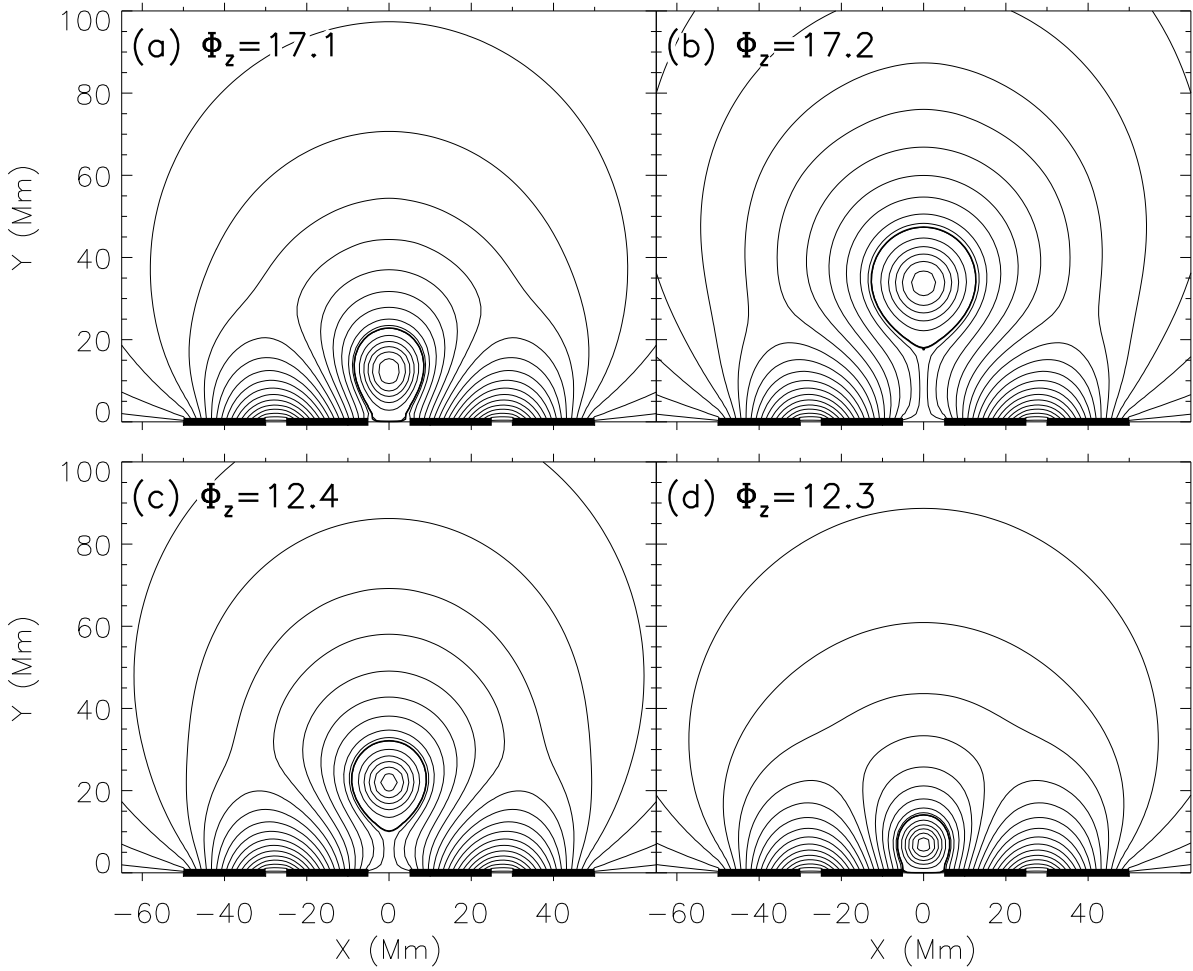}
\caption{Magnetic configurations of the flux rope system (a) right before and (b) after the upward catastrophe, (c) right before and (d) after the downward catastrophe for the case with $d_s=10.0$ Mm and $\sigma=1.0$.}\label{fig:evo}
\end{figure*}

The initial configurations with $d_s = 0.0, 2.0, 4.0, 6.0, 8.0, 10.0$ Mm are shown in \fig{fig:initd}(a)-\ref{fig:initd}(c) and \ref{fig:initd}(g)-\ref{fig:initd}(i), respectively, with the same $\sigma=1.0$. The photospheric distribution of the corresponding normal component of the magnetic field, $B_y$, is plotted in \fig{fig:initd}(d)-\ref{fig:initd}(f) and \ref{fig:initd}(j)-\ref{fig:initd}(l). $B_y$ corresponds to the radial component of the photospheric magnetic fields in observations. For each flux rope system, starting from the initial state, we increase the axial magnetic flux $\Phi_z$ to calculate different equilibrium solutions, as shown by the red dots in \fig{fig:fluxd}. The poloidal flux $\Phi_p$ of the rope for all equilibrium solutions in \fig{fig:fluxd} is fixed at $\Phi_p^0=-7.5\times10^9$ Mx cm$^{-1}$. \fig{fig:fluxd}(a)-\ref{fig:fluxd}(c) and \ref{fig:fluxd}(g)-\ref{fig:fluxd}(i) plot the evolutions of $H$, and \fig{fig:fluxd}(d)-\ref{fig:fluxd}(f) and \ref{fig:fluxd}(j)-\ref{fig:fluxd}(l) plot those of $L_c$. With increasing $\Phi_z$, the flux rope evolves from the sticky state to the levitating state. As shown in \fig{fig:fluxd}, the transition between these two kinds of states is quite different for different values of $d_s$. For the cases with small enough $d_s$, i.e. $d_s=0.0$ and $2.0$ Mm, the transition from the sticky state to the levitating state is continuous, indicating that these magnetic systems are non-catastrophic, whereas for $d_s\geqslant4.0$ Mm, the sticky and levitating equilibrium states are diverged into upper and lower branches, respectively, and the transition is manifested as a discontinuous jump from the lower branch to the upper branch, i.e. this is an upward catastrophe. Note that for the non-catastrophic cases, H and $L_c$ will saturates for further increasing $\Phi_z$, so that these flux rope systems should be non-eruptive. The upward catastrophic points are marked by the red vertical dotted lines in \fig{fig:fluxd}. An example of the upward catastrophe is exhibited in \fig{fig:evo}(a) and \ref{fig:evo}(b), which illustrate the equilibrium states of the systems with $d_s=10.0$ Mm just before and after the upward catastrophe. The flux rope keeps sticking to the photosphere till the upward catastrophic point $\Phi_z^u=17.2\times10^{18}$ Mx, across which the flux rope quickly jumps upward and levitates in the corona. The transition from BPSS configuration to HFT configuration with increasing axial flux has also been shown by studies based on NLFFF extrapolations \citep{Savcheva2009a,Su2011a,Savcheva2015a}. The simulation results reveal that, under force-free conditions, if the surface charges are not too close, upward catastrophe could also exists in quadrupolar background field.
\par
In the simulations discussed above (as shown by the red dots), we have obtained equilibrium states with the flux rope levitating in the corona for magnetic systems with different photospheric flux distributions. Starting from these levitating states, we decrease the control parameter $\Phi_z$ to calculate new equilibrium states, so as to investigate the transition from the levitating state to the sticky state following a route distinctive from the former transition. As shown by the blue triangles in \fig{fig:fluxd}, the type of this transition also varies with photospheric flux distributions. For the systems that have upward catastrophe, i.e. $d_s\geqslant4.0$ Mm, there also exists a downward catastrophe from the upper branch to the lower branch; the downward catastrophic points are marked by the blue dotted lines in \fig{fig:fluxd}. These cases are similar as that simulated in \cite{Zhang2016a}. The equilibrium solutions of the systems with $d_s=10.0$ Mm just before and after the downward catastrophe are also illustrated in \fig{fig:evo}(c) and \ref{fig:evo}(d); the downward catastrophic point is $\Phi_z^d=12.3\times10^{18}$ Mx. For the systems with $d_s=0.0$ and $2.0$ Mm, however, the transition from the levitating states to the sticky states is still continuous, indicating that there is no downward catastrophe either. Thus we may conclude that downward catastrophe also exists in quadrupolar background field under force-free conditions, and that the equilibrium states of the system are diverged into upper and lower branches by the upward and downward catastrophes when the photospheric flux is not concentrated too much towards the PIL, otherwise the transition between the sticky and levitating states is continuous so that neither upward nor downward catastrophe occurs. 
\par
Photospheric flux distribution also influences the properties of upward and downward catastrophes, as tabulated in \tbl{tbl:d}. The upward catastrophic point $\Phi_z^u$ increases with increasing $d_s$, which implies that the background field with a larger $d_s$ exerts a stronger constraint on the flux rope. The amplitude of the upward catastrophe $L_z^u$ also increases with increasing $d_s$, indicating that the upward catastrophe is more drastic in the system with larger $d_s$, namely, the system with larger $d_s$ trends to produce larger activities. The downward catastrophic point $\Phi_z^d$ is almost the same in catastrophic systems with different $d_s$, from which we may infer that the influence of $d_s$ on downward catastrophe might be somewhat different from that on upward catastrophe. This might be the probable reason for that the variation of the downward catastrophic amplitude $L_z^d$ with $d_s$ is also slightly different from that of $L_z^u$. Moreover, as seen from \tbl{tbl:d}, the separation between the two catastrophic points, $\Phi_z^u-\Phi_z^d$, also increases with increasing $d_s$. This indicates that, with increasing $d_s$, the system first evolves from a non-catastrophic one to a catastrophic one, and then the two catastrophes are increasingly separated.
\par
\begin{figure*}
\includegraphics[width=\hsize]{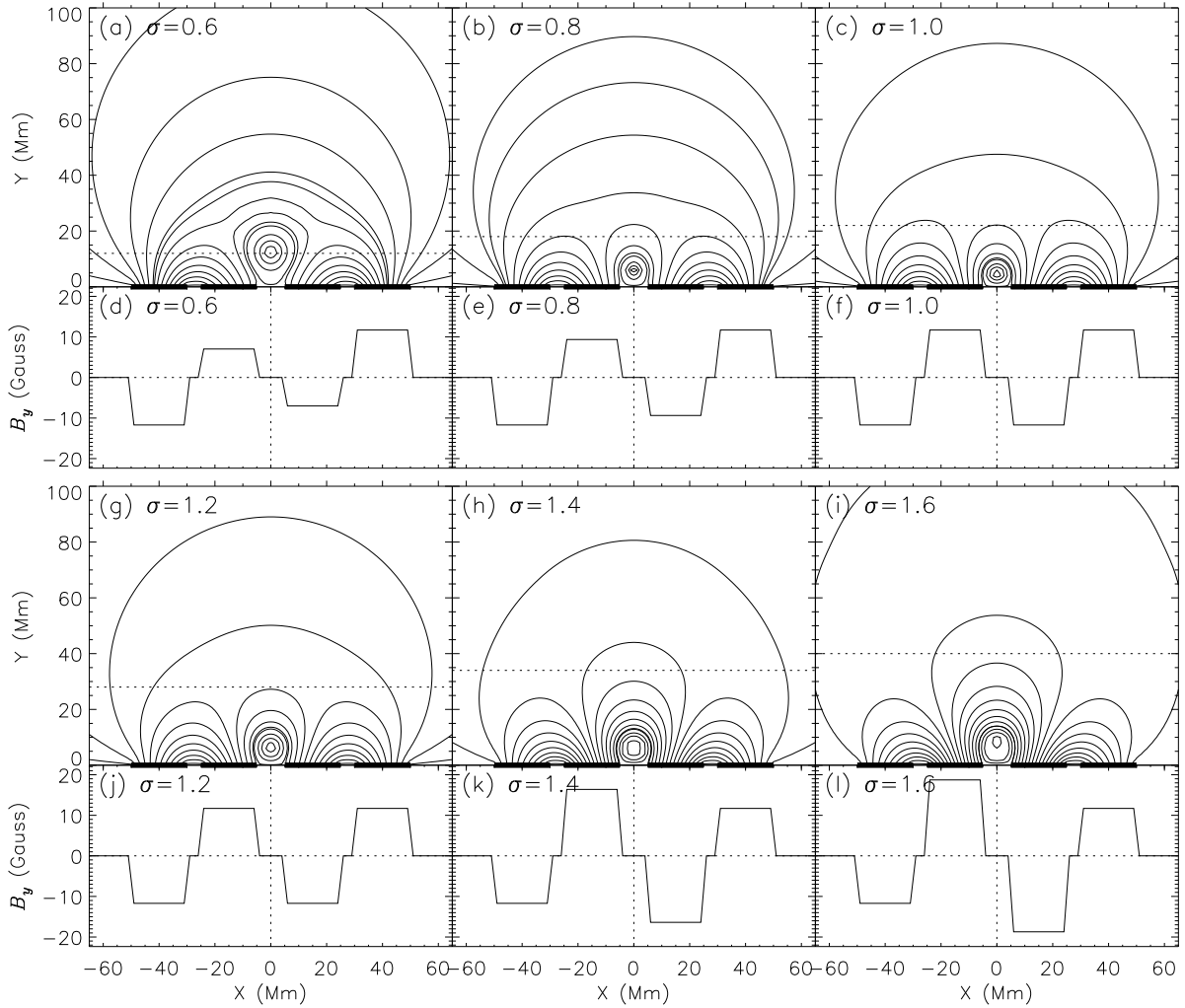}
\caption{The quadrupolar background configurations and the corresponding normal components of the magnetic field ($B_y$) at the photosphere ($y=0$) for different $\sigma$, which is selected to be $0.6, 0.8, 1.0, 1.2, 1.4, 1.6$ Mm, respectively; $d_s$ is 10.0 Mm for all the six cases. The two pairs of surface magnetic charges for different cases are marked by the black solid lines at $y=0$ in panels (a)-(c) and (g)-(i). The height of the neutral point in the quadrupolar background field is marked by the horizontal dotted line in panels (a)-(c) and (g)-(i).}\label{fig:initg}
\end{figure*}
\begin{figure*}
\includegraphics[width=\hsize]{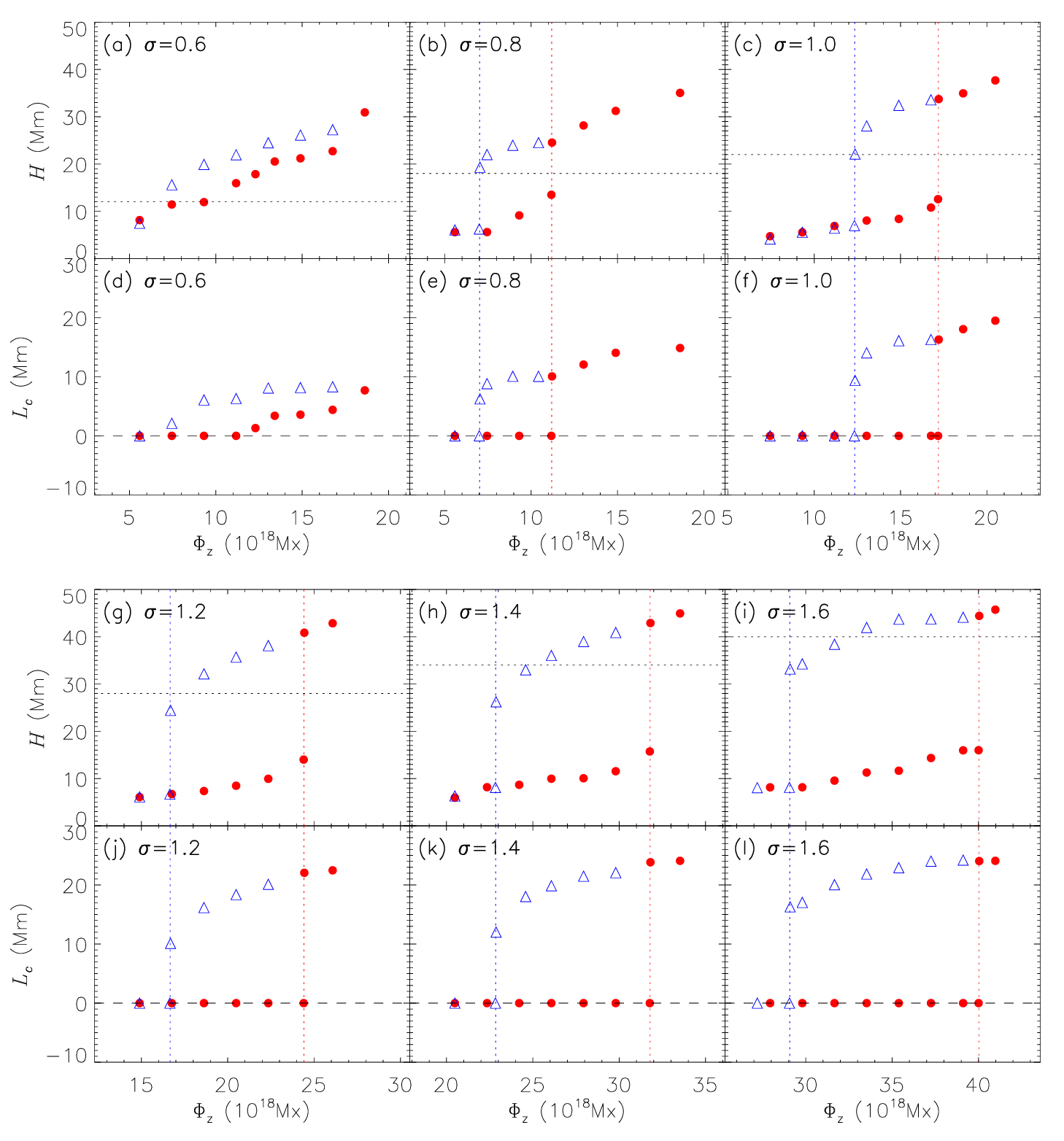}
\caption{$H$ and $L_c$ versus $\Phi_z$ for quadrupolar background fields with different $\sigma$ under force-free conditions. The meanings of the symbols are the same as those in \fig{fig:fluxd}.}\label{fig:fluxg}
\end{figure*}

Apart from $d_s$, we also adjust $\sigma$ through adjusting the charge density of the inner pair of surface charges (with fixed $d_s$) to obtain different photospheric flux distributions. The initial configurations with $\sigma = 0.6, 0.8, 1.0, 1.2, 1.4, 1.6$ are shown in \fig{fig:initg}(a)-\ref{fig:initg}(c) and \ref{fig:initg}(g)-\ref{fig:initg}(i), respectively, with the same $d_s=10.0$ Mm. The corresponding $B_y$ is plotted in \fig{fig:initg}(d)-\ref{fig:initg}(f) and \ref{fig:initg}(j)-\ref{fig:initg}(l). A smaller $\sigma$ implies a weaker inner pair of charges, corresponding to less magnetic flux of the background field above the flux rope, so that the constraint exerted by the background field on the flux rope is also weaker. Following similar simulating procedures as those introduced above, the evolution of the flux rope in systems with different $\sigma$ as a function of $\Phi_z$ are calculated, as shown in \fig{fig:fluxg}. The poloidal flux $\Phi_p$ of the rope is fixed at $-3.7\times10^9$ Mx cm$^{-1}$ for the case with $\sigma=0.6$, and at $-7.5\times10^9$ Mx cm$^{-1}$ for the other cases. For $\sigma=0.6$, the transition between the sticky state and the levitating state is always continuous, i.e. neither upward nor downward catastrophe exists in this system. Thus we may conclude that if the constraint of the background field on the flux rope is too weak, the system should be non-catastrophic. For $\sigma\geqslant0.8$, the equilibrium states of these systems are diverged into upward and downward branches, which are only connected by the upward and downward catastrophes. The properties of the catastrophes in magnetic systems with different $\sigma$ are tabulated in \tbl{tbl:g}. The catastrophic points of both the upward and downward catastrophes increase with increasing $\sigma$, so do the amplitudes of the catastrophes. This indicates that both the upward and downward catastrophes in the system with larger $\sigma$ are more drastic, so that the magnetic system with larger $\sigma$ trends to produce larger active region activities. The difference $\Phi_z^u-\Phi_z^d$ also increases with increasing $\sigma$, similar as that with $d_s$. 
\par
In summary, under force-free condition, the catastrophic behaviors of the magnetic system consisting of a flux rope in a fully closed quadrupolar background field are influenced by the photospheric magnetic conditions. The system could have both upward and downward catastrophe, provided that the photospheric flux distribution is not concentrated too much towards the PIL and the constraint exerted by the background field on the flux rope is not too weak. A downward catastrophe is always accompanied by an upward catastrophe, so that the equilibrium states of the system are diverged into two branches by these two catastrophes. With increasing $d_s$ and $\sigma$, the flux rope activities in the system trends to be stronger.

\subsection{Non-force-free condition}
\label{sec:non-free}
\begin{figure*}
\includegraphics[width=\hsize]{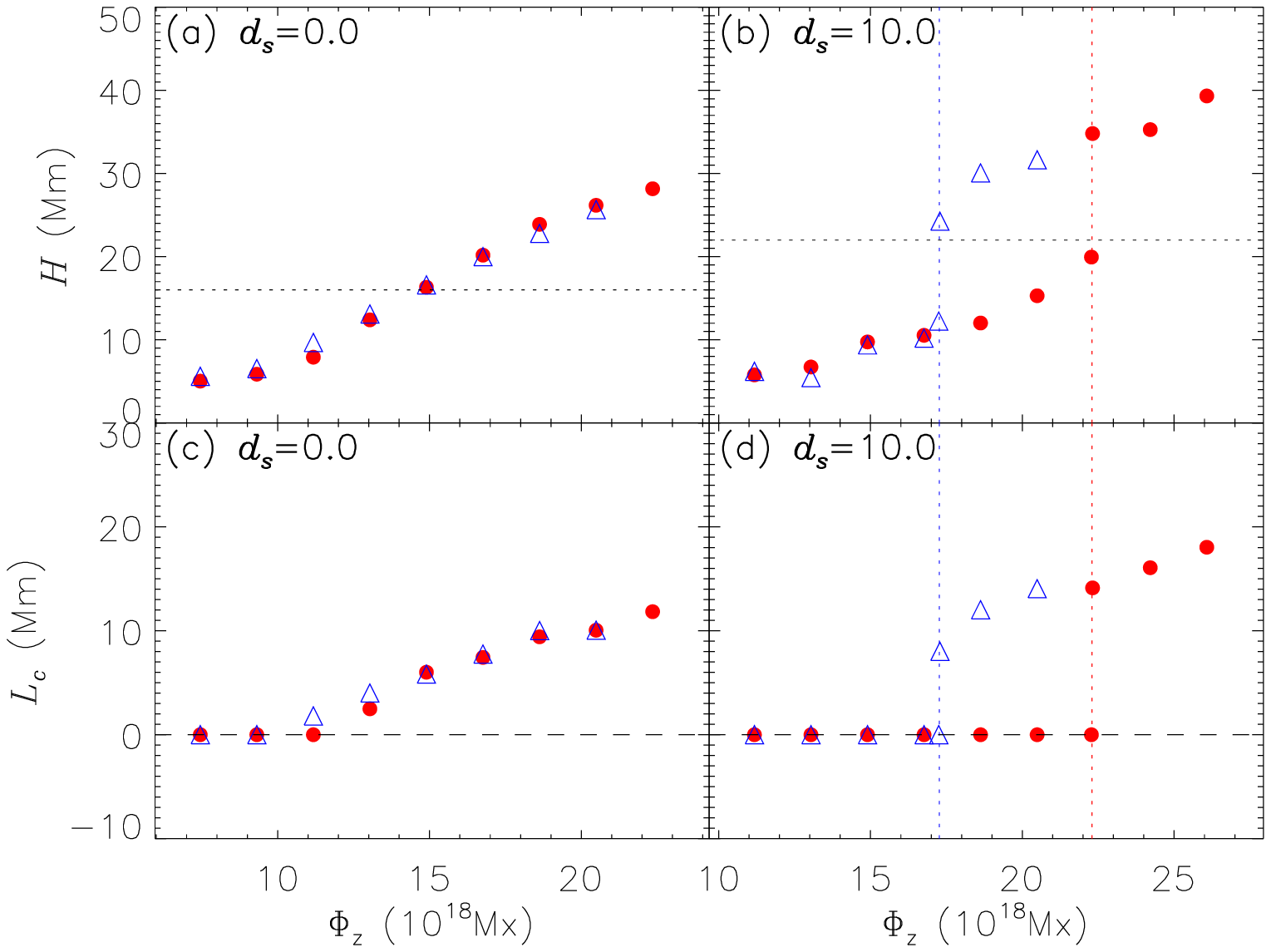}
\caption{$H$ and $L_c$ versus $\Phi_z$ for quadrupolar background fields with different $\sigma$ under non-force-free conditions. The meanings of the symbols are the same as those in \fig{fig:fluxd}.}\label{fig:fluxn}
\end{figure*}

Flux rope system does not always satisfy force-free approximation. For example, prominences are cool and dense plasma suspended in hot and diluted corona \citep[e.g.,][]{Liu2012a,Wang2010a,Liu2012b}, so that flux rope system containing a prominence should be far from force-free. To be comprehensive in our investigation of the evolution of a flux rope in a quadrupolar background field, we also simulate the evolution of the flux rope system under non-force-free condition. Here we calculate two cases: $d_s=0.0$ Mm, $\sigma=1.0$ and $d_s=10.0$ Mm, $\sigma=1.0$. Under non-force-free condition, the flux rope is characterised by not only the magnetic parameters, $\Phi_z$ and $\Phi_p$, but also $M$, the mass of the rope per unit length, which is always fixed at $M=334$ g cm$^{-1}$ in the simulation. Assuming that the length of the flux rope is about $100$ Mm, the mass of the flux rope would be $3.3\times10^{12}$ g, which is comparable to the lower values of the observed mass range of solar prominences \citep{Labrosse2010a,Parenti2014a}. Removing the relaxation procedure introduced in \sect{sec:equation}, the magneto-static equilibrium solutions with different $\Phi_z$ but the same $\Phi_p=-3.7\times10^9$ Mx cm$^{-1}$ are calculated. As shown in \fig{fig:fluxn}, when $d_s$ is large enough, there are also upward and downward catastrophes in quadrupolar field under non-force-free conditions, otherwise the geometric parameters vary continuously with increasing or decreasing $\Phi_z$. Thus we may conclude that upward and downward catastrophe also exist in quadrupolar field under non-force-free condition, and the existence of the catastrophes is influenced by the photospheric magnetic conditions, which are similar to the conclusions reached above under force-free condition.

\section{Upward catastrophe versus Torus instability}
\label{sec:torus}
The equilibrium of a coronal magnetic flux rope is usually simplified as the balance between the upward Lorentz force resulting from the oppositely directed image current of the flux rope (also called hoop force in some papers), and the downward Lorentz force from the constraint of the external poloidal magnetic field \citep{Kliem2014a}. By analyzing these two Lorentz forces acting on the flux rope, it is found that if the external magnetic field of a flux rope system, $B_{ex}$, decreases fast enough with the height above the photosphere, the flux rope is unstable to an upward disturbance, which is called ``Torus instability" \citep{Kliem2006a,Zuccarello2016a}. The decrease of the external field is described by the decay index $n=-\mathrm{d}(\mathrm{ln}B_{ex})/\mathrm{d}(\mathrm{ln}h)$. Based on wire current model, it is derived that torus instability occurs if $n$ is larger than 1 for straight current channels \citep{vanTend1978a,Filippov2001a} and 1.5 for circular cases \citep{Kliem2006a}. Both theoretical and observational studies found that torus instability plays an important role in triggering flux rope eruptions \citep[e.g.,][]{Torok2007a,Guo2010b}. 
\par
Catastrophes has close relationship with instabilities. By setting the analysis of loss of equilibrium and stability analysis in the same analytical framework, \cite{Demoulin2010a} suggested that upward catastrophe and torus instability should be two different views of the same physical mechanism. Furthermore, \cite{Kliem2014a} made a comprehensive analytical study about the relationship between the torus instability and the upward catastrophe triggered by the variations of the photospheric flux distributions, and found that the nose point of the equilibrium manifold, at which the upward catastrophe occurs, just connects the stable and unstable branches of the equilibrium states. In other words, at this nose point, not only upward catastrophe occurs, but also the system evolves from stable equilibrium to unstable equilibrium, so that torus instability occurs as well. Therefore, \cite{Kliem2014a} concluded that upward catastrophe and torus instability should be equivalent descriptions for the onset condition of solar eruptions.
\par
\begin{figure*}
\includegraphics[width=\hsize]{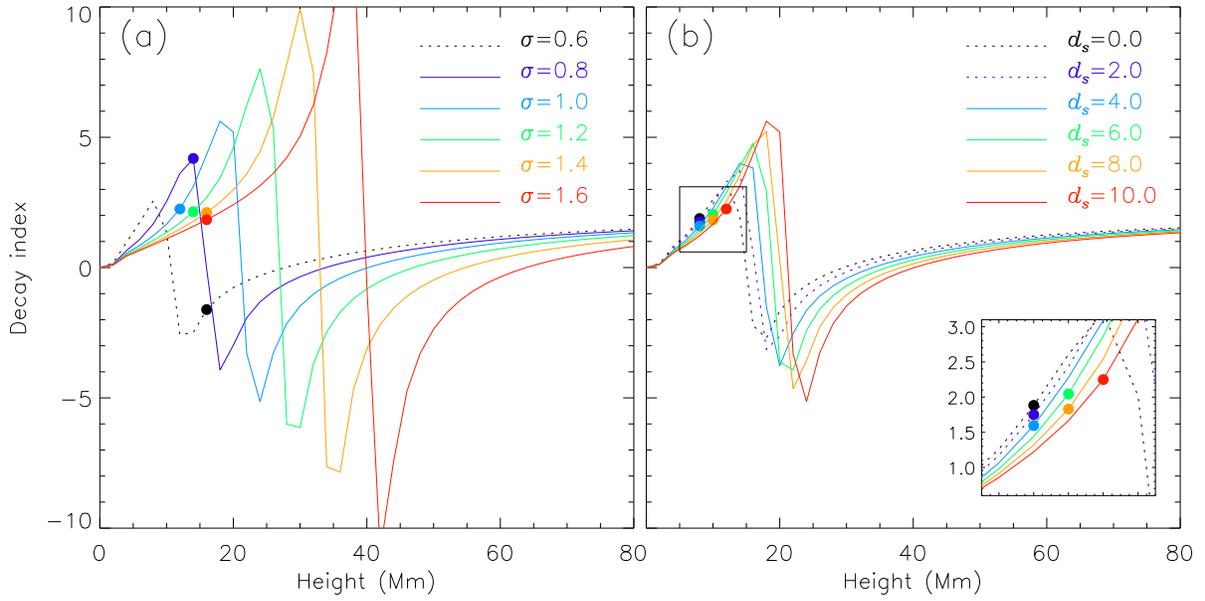}
\caption{Decay indices of the external field in different flux rope systems. Panel (a) shows the variations of the decay index with height along $x=0$ for different $\sigma$, and panel (b) shows those for different $d_s$; the catastrophic cases are plotted in solid lines, and the non-catastrophic ones in dotted lines.}\label{fig:decay}
\end{figure*}

In this paper, we have simulated the evolutions of the flux ropes systems versus the variation of the flux rope itself under different photospheric magnetic conditions. In order to investigate the role that torus instability plays in our simulation, we calculate the decay index of the external magnetic field under different photospheric magnetic conditions. Here the external magnetic field is just the background field for each case, which is a potential quadrupolar field. Note that the flux rope model in our simulation is different from that for torus instability in many aspects, so that the analysis here is only semiquantitative. \fig{fig:decay}(a) illustrates the variations of the decay index along $x=0$ for different $\sigma$. Non-catastrophic cases are plotted in dotted lines, and catastrophic ones in solid lines. The dot represents the location of rope axis in the equilibrium state right before the flux rope breaks away from the photosphere. It should be noted that our simulation is 2.5 dimensional, indicating that what we analyze here is the torus instability of a straight flux rope in quadrupolar field. As shown by \fig{fig:decay}(a), the decay indices at the rope axis for the catastrophic cases are at least 1.8, so that the flux rope is probably torus unstable in these catastrophic cases. For the non-catastrophic case with $\sigma=0.6$, however, the decay index at the rope axis is only -1.6, indicating a torus stable system. Therefore, for different flux rope systems with different $\sigma$, upward catastrophe is in good correspondence with torus instability. 
\par
\begin{figure*}
\includegraphics[width=\hsize]{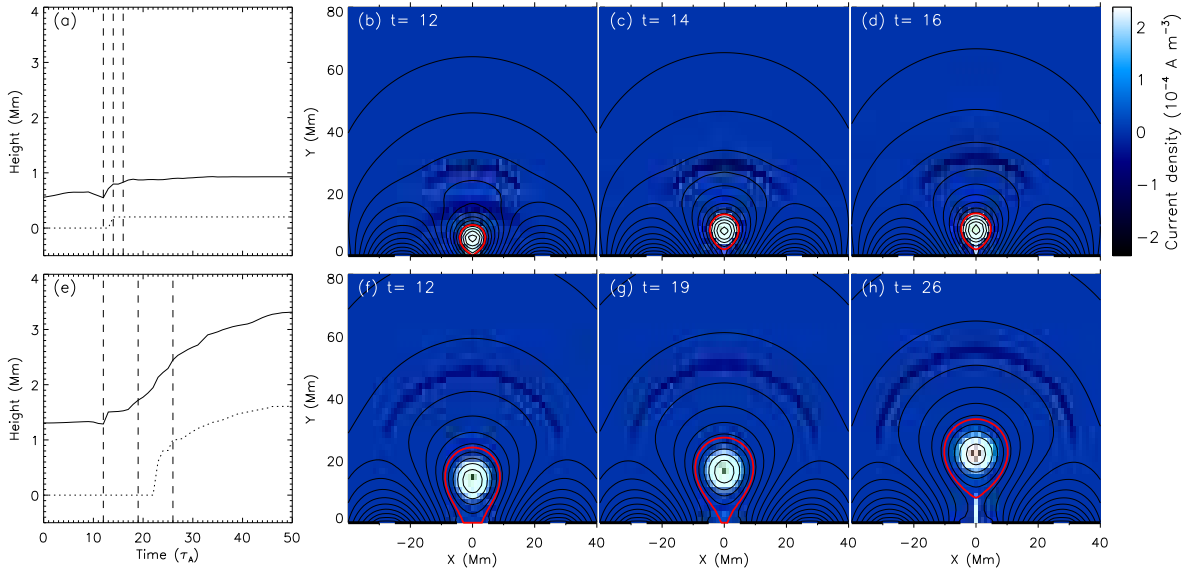}
\caption{Dynamic processes during the flux rope breaks away from the photosphere in non-catastrophic (top panels) and catastrophic (bottom) panels. Panel (a) plots the variations of the height of the rope axis (solid lines) and length of the current sheet below the flux rope (dotted lines) in the system with $d_s=0.0$ Mm. The distributions of $j_z$ at different times are illustrated in panels (b)-(d), as marked by the vertical dashed lines in panel (a), respectively. Panel (e) plots the variations of geometric parameters in the system with $d_s=10.0$ Mm, and the corresponding distributions of $j_z$ are illustrated in panels (f)-(h). The boundaries of the flux rope are marked by the red curves.}\label{fig:current}
\end{figure*}

The variations of the decay index along $x=0$ for different $d_s$ are shown in \fig{fig:decay}(b). Similarly, catastrophic and non-catastrophic cases are plotted in solid and dotted lines, respectively. Different from $\sigma$, all the flux rope systems with different $d_s$ have decay indices no less than 1.59, i.e. all these flux rope systems are probably torus unstable. This indicates that torus unstable systems could also be non-catastrophic. In order to find out the cause, we compare the dynamic processes during which the flux rope breaks away from the photosphere in non-catastrophic and catastrophic cases, as shown in \fig{fig:current}. \fig{fig:current}(a) is the calculation for the equilibrium state with $\Phi_z=11.4\times10^{18}$ Mx in the system with $d_s=0.0$ Mm, i.e. the state right after the flux rope breaks away from the photosphere. The variations of the height of the rope axis and the length of the current sheet below the flux rope are plotted by solid and dotted lines, respectively. The unit of time is $\tau_A=17.4$ s. \fig{fig:current}(b)-(d) are the distributions of the current in $z-$direction, $j_z$, at different times, as marked by vertical dashed lines in \fig{fig:current}(a); the boundary of the flux rope is marked by the red curves. At first, the flux rope sticks to the photosphere. Then adjust $\Phi_z$ to 11.4$\times10^{18}$ Mx and let the system relax to equilibrium state. At the time $t=12~\tau_A$, the flux rope begins to break away from the photosphere (\fig{fig:current}(b)), which is almost immediately followed by the appearance of the current sheet below the flux rope at $t=14~\tau_A$ (\fig{fig:current}(c)). Since the net current within the flux rope and the current sheet below the flux rope have the same direction, a downward force is exerted by the newly-formed current sheet on the flux rope, so that the upward motion of the flux rope is also immediately terminated (\fig{fig:current}(d)), resulting in a continuous transition from the sticky state to the levitating state. Although the decay index is rather large (here $n=$1.9), torus instability should be prohibited by the quickly generated current sheet below the flux rope at the very beginning (B. Kliem, private communication). Then the force balance of the flux rope is determined by not only the hoop force and the Lorentz force of the external field, but also the drag force from the current sheet below the rope. If $\Phi_z$ is larger, stronger drag force is also needed, so that the flux rope levitates at a higher height with a longer current sheet beneath. Therefore, in the system with $d_s=0.0$ Mm, the $H$ and $L_c$ increases continuously with the control parameter $\Phi_z$, so that the system is non-catastrophic. The system with $d_s=0.1$ Mm also has similar conclusion. \fig{fig:current}(e) is the calculation for the equilibrium state with $\Phi_z=17.2\times10^{18}$ Mx in the system with $d_s=10.0$ Mm, i.e. the state right after upward catastrophe occurs; \fig{fig:current}(f)-(h) are the corresponding distributions of $j_z$. Different from that in \fig{fig:current}(a), although the flux rope also begins to break away from the photosphere at $t=12~\tau_A$, the current sheet below the flux rope does not appears until $t=22~\tau_A$, as shown in \fig{fig:current}(g). During this period, the underside of the flux rope still keeps sticking to the photosphere (see \fig{fig:current}(g)), which is somewhat similar as the line-tied effect \cite[e.g.,][]{Isenberg2007a,Aulanier2010a}. Since there is no current sheet below the flux rope during this period, and torus unstable condition should be satisfied (here $n$=2.2), torus instability occurs so that the flux rope keeps rising. After $t=22~\tau_A$, the flux rope detaches from the photosphere (see \fig{fig:current}(h)), and the current sheet appears below the flux rope, which prevents the further evolution of torus instability. Eventually, the flux rope levitates at a certain height, resulting in a equilibrium state discontinuous from the state with $\Phi_z<17.2\times10^{18}$ Mx, whose $L_c$ is always 0. Therefore there is an upward catastrophe in the system with $d_s=$10.0 Mm. The major difference in this system from that with $d_s=$0.0 Mm is the obvious delay of the appearance of current sheet, during which torus instability could evolves to certain extent, so that upward catastrophe could occur in this system.
\par
In summary, we may infer that torus unstable might be a necessary but not sufficient condition for upward catastrophe; torus unstable systems could also be non-catastrophic. This is because the analysis for torus instability does not take the effect of the current sheet below the flux rope into account. Our simulation results demonstrate that the current sheet below the flux rope is also important for the onset condition of solar eruption: if the current sheet appears immediately after the flux rope move upwards, torus instability will be prohibited at the very beginning, so that the system is non-catastrophic, and as a result, there is no eruption in this system. As discussed above, the appearing time of the current sheet is significantly influenced by photospheric flux distributions.
\par

\section{Discussion and Conclusion}
\label{sec:discuss}
To investigate the catastrophic behavior of flux rope systems in strong active regions, we simulate the evolution of the magnetic system consisting of a flux rope in fully closed quadrupolar background fields with different photospheric flux distributions. Under force-free condition, it is found that, when the photospheric flux is not concentrated too much towards the PIL (large enough $d_s$) and the constraint exerted by the background field is not too weak (large enough $\sigma$), the equilibrium states of the system are separated into two branches, which are connected by an upward and a downward catastrophe, respectively. Otherwise, the geometric parameters always evolve continuously with varying $\Phi_z$. Therefore, we may conclude that downward catastrophe also exists in quadrupolar fields, and the upward and downward catastrophes are always paired with each other. Moreover, the properties of both the upward and the downward catastrophes are also influenced by the photospheric flux distribution; larger $d_s$ and $\sigma$ not only favors the existence of the catastrophes, but also result in more drastic evolutionary profile when there exist catastrophes, namely, a system with larger $d_s$ and $\sigma$ trends to produce stronger active region activities. Similar conclusion also holds for the magnetic system under non-force-free condition. The magnetic configuration in our simulation is similar as that in the breakout scenario, in which the eruption is triggered by the reconnection at the upper current sheet. By simulations in bipolar field, \cite{Zhang2017a} found that, catastrophe only exists when the photospheric flux is concentrated not too much towards the central region and the background field is not too weak, which is consistent with our simulation results in quadrupolar field. 
\par
For the flux rope systems with different photospheric magnetic conditions, we also calculate the decay index at the rope axis in the state right before the flux rope leaves the photosphere. It is revealed that upward catastrophe and torus instability should have close relationship: catastrophic flux rope systems trend be torus unstable, whereas torus unstable systems may not always be catastrophic; the current sheet below the flux rope might also be important for the onset of flux rope eruptions. In our simulation, the flux rope has a finite cross section, so that the critical decay index derived based on wire current model, 1 for straight current channel and 1.5 for circular current channel, could hardly be directly used in the analysis for our simulation results, so that the analysis here is only semiquantitative. Since a downward catastrophe is always paired with an upward catastrophe in our simulation, we may infer that the non-eruptive downward catastrophe also tends to occur in the magnetic system with strong decay of the magnetic fields above the flux rope.
\par
By using a simpliﬁed analytic flux rope model in quadrupolar magnetic fields, \cite{Longcope2014a} analyzed the quasi-static evolution with the changes at the boundary or the reconnection above the flux rope and under it, and found that all these three kinds of evolutionary scenarios can lead to catastrophe. In our simulations, it is demonstrated that the processes resulting in the changes of the properties in the flux rope can also trigger catastrophes, and both the existence and properties of the catastrophes are influenced by the photospheric magnetic conditions. The analytical study in \cite{Longcope2014a} and our simulations reveal different aspects of the catastrophes in quadrupolar magnetic fields.
\par

\par
We are grateful to Dr. Bernhard Kliem for his guidance and suggestions in the analysis about the relationship between catastrophe and torus instability. We also appreciate the anonymous referee for his/her valuable comments that significantly improved this paper. This research is supported by Grants from NSFC 41131065, 41574165, 41421063, 41474151 and 41222031, MOEC 20113402110001, CAS Key Research Program KZZD-EW-01-4, and the fundamental research funds for the central universities WK2080000077. R.L. acknowledges the support from the Thousand Young Talents Program of China.


\begin{deluxetable}{cccccccc}
\tablewidth{0pt}
\tablecaption{Parameters of the catastrophes versus $d_s$ for $\sigma=1.0$\label{tbl:d}}
\tablehead{ \colhead{$d_s$(Mm)} & \colhead{$\Phi_z^u$($10^{10}$ Wb)} & \colhead{$L_c^u$(Mm)} & \colhead{$\Phi_z^d$($10^{10}$ Wb)} & \colhead{$L_c^d$(Mm)} & \colhead{$\Phi_z^u-\Phi_z^d$($10^{10}$ Wb)}}
\startdata
4.0  & 14.4  & 10.0  & 12.1  & 8.0   & 2.3\\
6.0  & 15.7  & 12.4  & 12.4  & 6.0   & 3.3\\
8.0  & 15.9  & 14.0  & 12.4  & 6.0   & 3.5\\
10.0 & 17.2  & 16.3  & 12.3  & 9.4   & 4.9
\enddata
\tablecomments{$\Phi_z^u$ and $\Phi_z^d$ represent the upward and downward catastrophic points, respectively; $L_c^u$ and $L_c^d$ are the spatial amplitudes of the upward and downward catastrophes, respectively.}
\end{deluxetable}

\begin{deluxetable}{cccccc}
\tablewidth{0pt}
\tablecaption{Parameters of the catastrophes versus $\sigma$ for $d_s=10.0$ Mm\label{tbl:g}}
\tablehead{ \colhead{$\sigma$} & \colhead{$\Phi_z^u$($10^{10}$ Wb)} & \colhead{$L_c^u$(Mm)} & \colhead{$\Phi_z^d$($10^{10}$ Wb)} & \colhead{$L_c^d$(Mm)} & \colhead{$\Phi_z^u-\Phi_z^d$($10^{10}$ Wb)}}
\startdata
0.8  & 11.2  & 10.1  &  7.0  &  6.3  & 4.2\\
1.0  & 17.2  & 16.3  & 12.3  &  9.4  & 4.9\\
1.2  & 24.4  & 22.0  & 16.6  & 10.1  & 7.8\\
1.4  & 31.8  & 23.8  & 22.8  & 12.0  & 9.0\\
1.6  & 40.0  & 24.0  & 29.0  & 16.3  & 11.0
\enddata
\tablecomments{The meanings of the parameters are the same as those in \tbl{tbl:d}.}
\end{deluxetable}

\end{document}